\documentclass[aps,prl,twocolumn,superscriptaddress,groupedaddress]{revtex4}
\usepackage{graphicx}
\usepackage{dcolumn} 
\usepackage{bm}   
\usepackage{color}
\usepackage{amssymb} 
\usepackage{natbib} 

\begin{document}
\setlength\parindent{5pt}
\widetext
\title{Changes of Fermi Surface Topology due to the Rhombohedral Distortion in  SnTe  }
\author{Christopher D. O'Neill}
\affiliation{Centre for Science at Extreme Conditions and SUPA, School of Physics and Astronomy, University of Edinburgh, Edinburgh, EH9 3JZ, UK}
\author{Oliver J. Clark}
\affiliation{School of Physics and Astronomy, University of St Andrews, St Andrews KY16 9SS, United Kingdom}
\author{Harry D.J. Keen}
\affiliation{Centre for Science at Extreme Conditions and SUPA, School of Physics and Astronomy, University of Edinburgh, Edinburgh, EH9 3JZ, UK}
\author{Federico Mazzola} 
\affiliation{School of Physics and Astronomy, University of St Andrews, St Andrews KY16 9SS, United Kingdom}
\author{Igor Markovi\'{c}}
\affiliation{School of Physics and Astronomy, University of St Andrews, St Andrews KY16 9SS, United Kingdom}
\affiliation{Max Planck Institute for Chemical Physics of Solids, N\"{o}thnitzer Stra\ss e 40, 01187 Dresden, Germany}
\author{Dmitry A. Sokolov}
\affiliation{Max Planck Institute for Chemical Physics of Solids, N\"{o}thnitzer Stra\ss e 40, 01187 Dresden, Germany}
\author{Andreas Malekos}
\affiliation{Centre for Science at Extreme Conditions and SUPA, School of Physics and Astronomy, University of Edinburgh, Edinburgh, EH9 3JZ, UK}
\author{Phil D. C. King}
\affiliation{School of Physics and Astronomy, University of St Andrews, St Andrews KY16 9SS, United Kingdom}
\author{Andreas Hermann}
\affiliation{Centre for Science at Extreme Conditions and SUPA, School of Physics and Astronomy, University of Edinburgh, Edinburgh, EH9 3JZ, UK}
\author{Andrew D. Huxley}
\affiliation{Centre for Science at Extreme Conditions and SUPA, School of Physics and Astronomy, University of Edinburgh, Edinburgh, EH9 3JZ, UK}
\date{\today}
\begin{abstract}
\noindent  
\small
Stoichiometric SnTe is theoretically a small gap semiconductor that undergoes a ferroelectric  distortion  on cooling. In reality however, crystals are always non-stoichiometric and metallic; the ferroelectric transition is therefore more accurately described as a polar structural transition.    Here we study the Fermi surface using quantum  oscillations  as a function of pressure. We find the oscillation spectrum changes at high pressure, due to the suppression of the polar transition and  less than 10~kbar is sufficient to stabilize the undistorted cubic lattice.  This is accompanied by a large decrease in the Hall and electrical resistivity. Combined  with our density functional theory (DFT) calculations and angle resolved photoemission spectroscopy (ARPES) measurements this suggests the Fermi surface $L$-pockets have lower mobility than the  tubular Fermi surfaces that connect them. Also captured in our DFT calculations is a small widening of the band gap and shift in density of states for the polar phase.  Additionally we find the unusual phenomenon of a  linear magnetoresistance that exists irrespective of the distortion that we attribute to regions of the Fermi surface with high curvature.          \end{abstract}
\maketitle
On first appearance SnTe looks like  a simple  semiconductor with a small bandgap that is formed by the ionic transfer between Sn and Te.  Despite its simplicity SnTe is known to host several interesting phenomena including ferroelectricity \cite{Pawley, ONeill} and is a  topological crystalline insulator with potential surface states  \cite{Hsieh, Tanaka}. Grown samples are however always  non-stoichiometric and Te rich, with the Sn deficiency accommodated in the lattice as vacancies  \cite{Brebrick}.  As a consequence  the Fermi level is shifted into the valence band, leading to metallic behaviour with  a  free carrier concentration of  holes, $n_h$.  The polar state transition temperature, $T_c$, is below 100~K and $n_h$ dependent \cite{Kobayashi}. Increasing $n_h$ reduces $T_c$ and eventually stabilizes the undistorted  lattice.  While avoiding the polar distortion is advantageous for studying  topological states which are protected by  crystalline symmetry, the  large $n_h$ makes any investigation of surface states by  transport extremely difficult. So far there are no reports on how $T_c$ varies under pressure.  Here we use hydrostatic pressure to suppress $T_c$ and measure the associated changes in  Fermi surface topology and transport properties. \\
\indent At 300 K SnTe has a $fcc$ rocksalt structure shown in Figure 1(a) with space group $Fm\overline{3}m$. The polar distortion is  driven by a soft  transverse optic phonon \cite{Pawley, ONeill},  that  distorts the lattice along the (111) direction of the cubic structure to a $R3m$ rhombohedral phase with shear angle $\alpha \approx 59.878^\circ$~\cite{Muldawer} ($\alpha=60^\circ$ for $fcc$) and relative shift of the two $fcc$ sublattices of $\upsilon \sim 0.008$ ($\sim 9$ pm)~\cite{Iizumi}.  Previous band structure calculations of  the cubic phase show a valence band made up of filled Te orbitals and a conduction band of empty Sn orbitals, except at the narrowest gap  (1,1,1) $L$-point of the Brillouin zone, where a spin-orbit driven band inversion occurs~\cite{Tung, Littlewood_Arpes}.  For small $n_h$ the cubic Fermi surface is made up of disconnected  pockets at the $L$-points. On increasing $n_h$ the pockets elongate and above $n_h\sim1\times10^{20}$cm$^{-3}$, are joined by tubes   \cite{ Littlewood_Arpes, Allgaier}.  Quantum oscillation measurements on bulk crystals resolved 3-4  frequencies belonging to the $L$-pockets for low values of $n_h$  \cite{Savage, Burke}.  These merged to a single frequency  at increased doping levels  where the undistorted lattice is anticipated to be stable at 0~K.  The distortion  therefore clearly has  consequences for the Fermi surface topology, that have yet to be fully determined.    \\
\indent   Angle resolved photoemission spectroscopy (ARPES) measurements could not resolve any changes in Fermi surface topology with temperature below $T_c$, however they did see a   shift in much of the density of states and a widening of the bandgap~\cite{Littlewood_Arpes}.  While the bandgap is a result of ionic transfer between Sn and Te, its widening  below $T_c$  suggests the polar transition has an electronic component resembling  a Peierls instability. Such a  mechanism would indeed lead to  changes in  Fermi-surface topology.  Recent measurements on thin films~\cite{Okazaki} also  suggest the possibility that the different quantum oscillation frequencies in the rhombohedral phase may be a result of  spin splitting by a spin-orbit Rashba mechanism~\cite{Zhang_2014}. \\
\indent  Avoiding $T_c$ using pressure on a sample with fixed doping  provides a unique route to investigate any changes in Fermi surface associated with the distortion.  Here we report   that a pressure of 10~kbar is adequate to suppress the distortion in a sample with $T_c=89$~K at 6~kbar pressure. \\
\indent Single crystals of SnTe were grown by the same method described elsewhere \cite{ONeill}  with high purity elements Sn (99.9999 $\%$) and Te (99.9999 $\%$) in a  ratio 51:49  to minimize $n_h$. Crystals had natural facet faces along cubic axes and were  orientated by Laue X-ray diffraction.  The $T_c$ and $n_h$ of the crystals  was determined from their electrical resistivity and Hall effect respectively.  A list of the samples studied  along with their $T_c$ and $n_h$ is given in Table I (S3 was used in pressure studies).   The values of $T_c$ are in close agreement with previous reports for samples with similar $n_h$ at ambient pressure in \cite{Kobayashi}. \\
\
 \begin{center}
\begin{tabular}{|c|c|c|}
\hline
 Sample (pressure)   &$T_c$ (K)   & $n_h (\times 10^{20}$ cm$^{-3}$)\\ \hline 
S1  (ambient)& 79  &$2.93\pm0.15$\\\hline
S2 (ambient)& 75&$3.09\pm0.14$\\\hline
S3 (6 kbar) & 89&$1.12\pm0.04$ \\\hline
\end{tabular}  
\end{center}
\noindent Table I : {\footnotesize Values of polar transition temperature, $T_c$, and corresponding carrier concentration, $n_h$, for 3 single crystal samples studied here.}\\

\indent   ARPES measurements were performed on sample S1 at the CASSIOPEE beamline of Synchrotron SOLEIL using a Scienta R4000 hemispherical analyser with a vertical entrance slit and light incident in the horizontal plane. The sample was cleaved \textit{in-situ} and measured at temperatures below 15~K. Measurements were taken with p-polarised light at photon energies of 110 and 135~eV. The approximate positions in $k_z$ of  $k_x$-$k_y$ Fermi surface contours and band dispersions were determined from the experimentally observed periodicity of band features as a function of photon energy in conjunction with reference density-functional theory calculations.\\
\indent Samples S2 and S3   were   cut by spark erosion into blocks with [001] cubic axes along the length, width and height and had dimensions  $1.80 \times 0.45 \times 0.32 $~mm and  $175 \times 115\times100$~$\mu$m respectively. Gold electrical contacts were attached by spot welding and  resistivity ($\rho$), Hall effect and magnetoresistance (MR) measurements made with a current of  100~$\mu$A at 37~Hz.  Experiments were carried out in a $^4$He cryostat with 9~T Cryogenic Ltd magnet and a dilution refrigerator with  15~T Oxford Instruments magnet.  Sample S2  was used to study quantum oscillations as a function of applied field angle with a rotating sample stage. Sample S3 was studied under pressure. Pressure was applied in  a  diamond anvil pressure cell that had diamond culets of  $800$~$\mu$m  diameter.   Daphne oil 7373 was used as the pressure transmitting medium. A  pre-indented steel gasket with a 300~$\mu$m hole,  insulated with  a Al$_2$O$_3$ and Stycast 1266 epoxy  mixture, contained the sample under pressure.      The pressure was determined via fluorescence lines of a small ruby chip at 300~K.   \\  
\begin{figure}[t]
\includegraphics[scale=0.6]{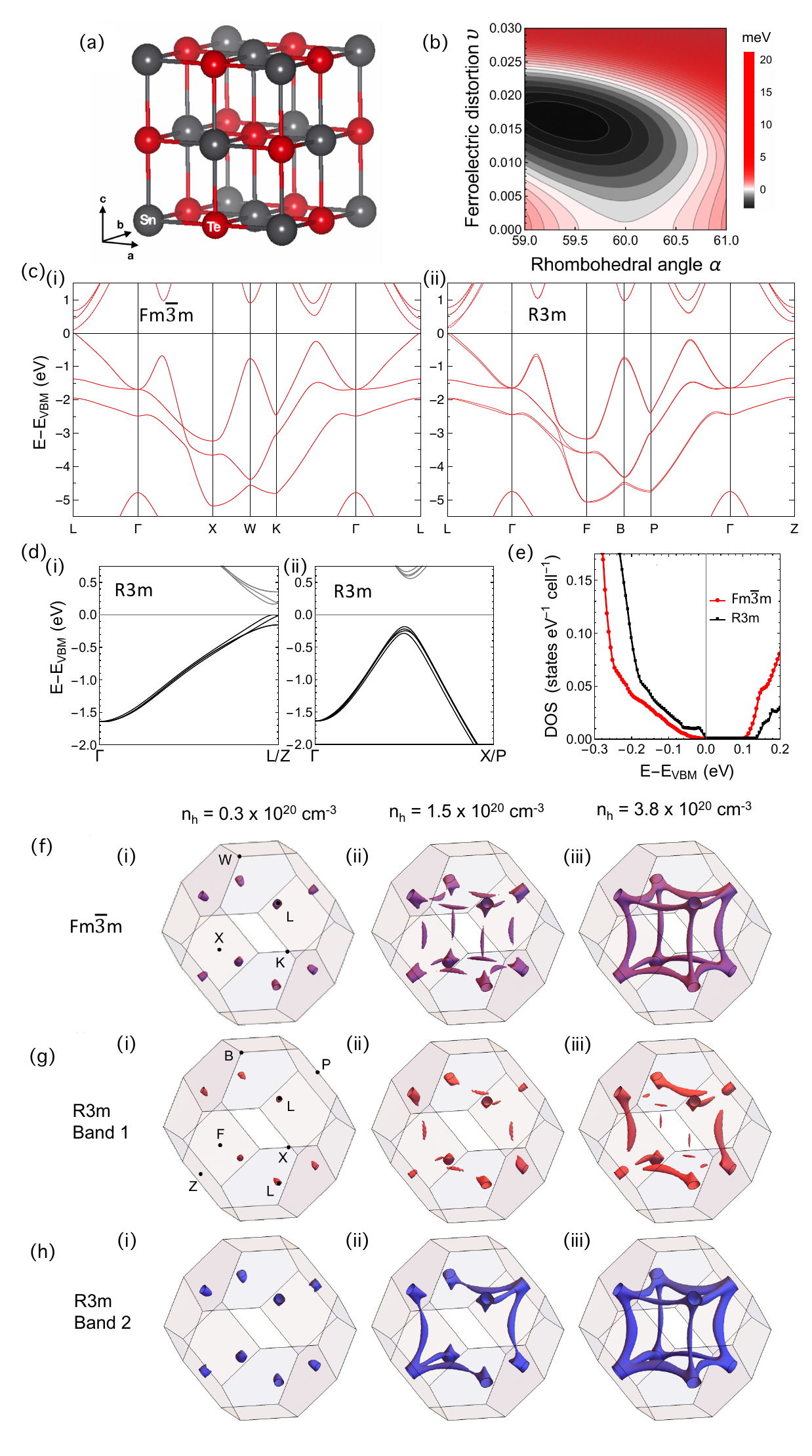}
\caption{\footnotesize (a) The cubic structure of SnTe at 300~K. (b)  The energetic landscape of the rhombohedral distortion at 0~K. Here the cubic structure is at zero energy and is a saddle point at $\alpha=60^\circ$ and $\upsilon=0$. (c)~The electronic band structure of stoichiometric SnTe along equivalent high symmetry paths in the (i) $Fm\overline{3}m$ cubic and (ii) $R3m$ rhombohedral structures. Here the band energy ($E$) is plotted relative to the valence band maximum ($E_\textrm{\tiny VBM}$).  (d) A close-up view of the valence band dispersion for the $R3m$ structure along directions relevant to the Fermi surface topology. Here degeneracy has been lifted along both (i) $\Gamma$-$L/Z$ and (ii) $\Gamma$-$X/P$  directions  as a result of the distortion. (e)~The electronic density of states as a function of $E-E_\textrm{\tiny VBM}$ for each structure. (f)~The evolution of the Fermi surface for the $Fm\overline{3}m$ structure at three positions of the Fermi energy ($E_F$) with respect to $E_\textrm{\tiny VBM}$; (i)~$-0.165$~eV, (ii)~$-0.282$~eV and (iii)~$-0.330$~eV. Corresponding carrier concentrations, $n_h$, are marked in the legend. (g)-(h)~The evolution of the Fermi surface for the  $R3m$ structure made up of 2 non-degenerate bands. Here the $E_F$ shifts upwards to (i)~$-0.128$, (ii)~$-0.232$ and (iii)~$-0.290$~eV, in order to conserve $n_h$ values.}
\end{figure} 
Fully relativistic density functional theory calculations were carried out for stoichiometric SnTe with the VASP package \cite{VASP.Kresse1996,PAW.Joubert1999,GGA.PBE1996} for a plane-wave basis with cutoff energy $E_c=240$ eV and $k$-point sampling with a linear density of 60/\AA$^{-1}$ (30/\AA$^{-1}$ for mapping the potential energy surface). Constant-pressure structure optimisations were performed until remaining force components were less than 1~meV/\AA. We found the ground state energy for the rhombohedral structure to be $\sim 2$~meV/unit lower than the cubic structure, with a rhombohedral angle distortion of $\alpha\approx 59.66^\circ$ and polar displacement of $\upsilon\approx 0.012$, somewhat larger than found in previous scalar relativistic calculations \cite{ONeill}. The potential energy surface for the distortion is shown in Figure 1(b). These values closely resemble those seen by X-ray and neutron scattering ($\alpha\approx 59.878^{\circ}$ and $\upsilon = 0.007$ for $n_h = 1.0$ and $0.88 \times 10^{20}$ cm$^{-3}$ respectively) \cite{Muldawer, Iizumi}. The calculated band structure in the $Fm\overline{3}m$ cubic phase is shown in Figure 1(c)(i) where the smallest band gap is seen at the $L$-point, as expected. The evolution of the Fermi surface with hole doping, within the frozen band approximation for this structure, is shown in Figure 1(f) with the Fermi energy ($E_F$) at (i) -0.165~eV ($n_h=0.3\times 10^{20}$~cm$^{-3}$), (ii) -0.282~eV ($n_h=1.5\times 10^{20}$~cm$^{-3}$) and (iii) -0.332~eV ($n_h=3.8\times 10^{20}$ cm$^{-3}$) with respect to the top of the valence band. Below $\sim-0.25$ eV (above $n_h\sim 1 \times 10^{20}$ cm$^{-3}$) the band along $\Gamma$-$K$ crosses  $E_F$ and ultimately leads to tubes connecting the $L$-pockets.\\
\indent The electronic band structure for the $R3m$ rhombohedral structure is shown in Figure 1(c)(ii). Here, the band gap at $L$ widens from 0.10 eV to 0.14 eV along with a shift in the density of states shown in Figure 1(e). Additionally, the breaking of cubic symmetry lifts degeneracies throughout the Brillouin zone. The paths relevant to the Fermi surface topology are $\Gamma$-$L/Z$ and $\Gamma$-$X/P$ ($\Gamma$-$L$ and $\Gamma$-$K$ in cubic notation),  shown in Figure 1(d)(i) and (ii) respectively. The resulting Fermi surface now consists of hole pockets from two, non-degenerate bands in Figure 1(g)-(h). The single eight-fold degenerate $L$-pockets from the cubic phase is split into four unique pockets, two per band. This is a consequence of both rhombohedral symmetry breaking, and spin-orbit splitting to lift the band degeneracy. The total carrier concentration (Fermi surface volume) must remain constant between the two phases. Hence, the $R3m$ phase sees an upward shift of $E_F$ to (i) -0.128~eV  (ii) -0.232 eV and (iii) -0.290 eV for the equivalent carrier concentrations in Figure 1(f).\\
 \begin{figure}[t]
\includegraphics[scale=0.71]{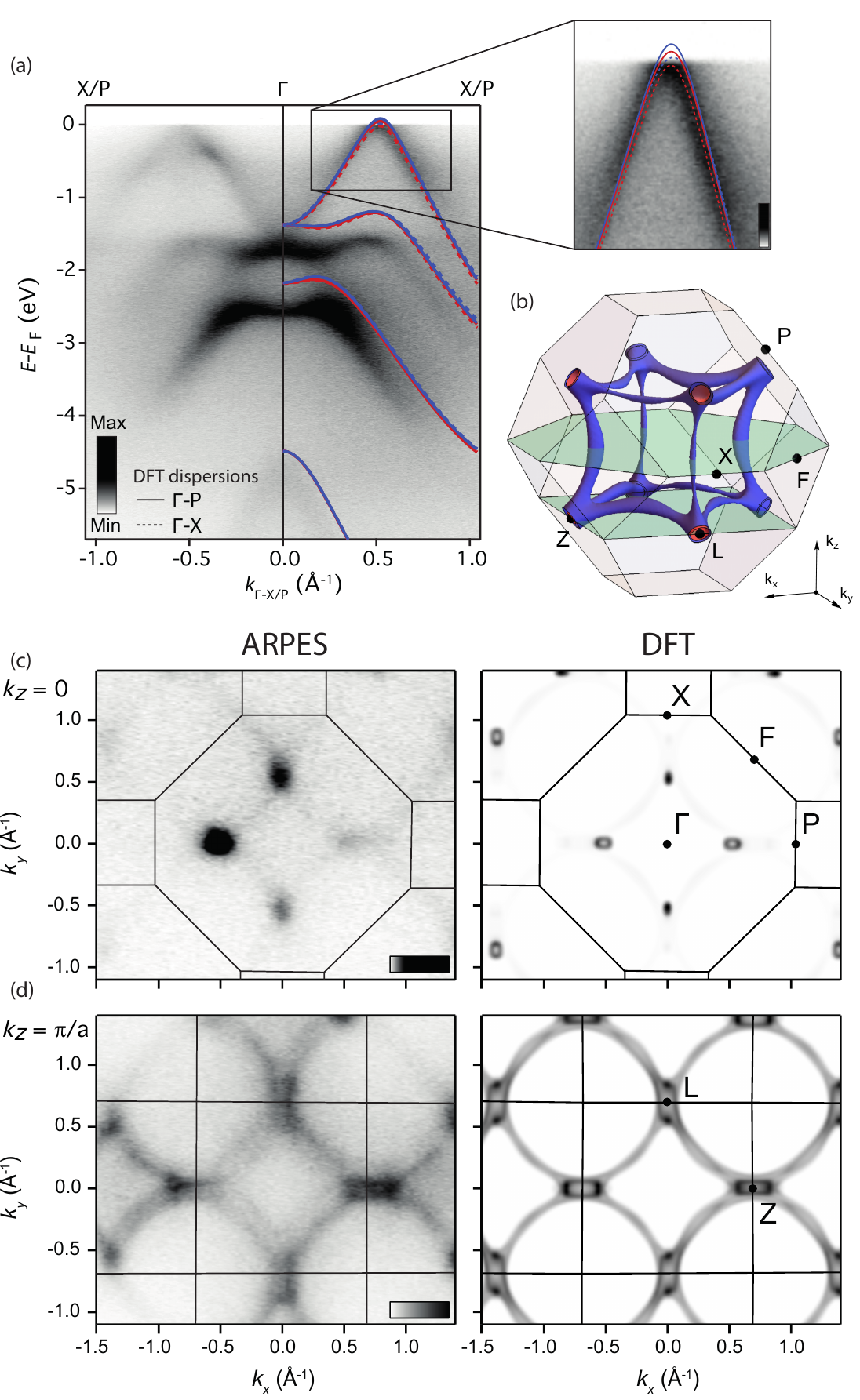}
\caption{\footnotesize  (a) Angle-resolved photoemission (ARPES) band dispersions along the $X/P$-$\Gamma$-$X/P$ direction measured with h$\nu=135$~eV light (grey scale). Density functional theory (DFT) calculations for  the  $R3m$ structure, shifted in energy  by 0.27 eV, are included as overlays (lines). (b)~Three-dimensional schematic of the $R3m$ Brillouin zone with the calculated Fermi surface for a Fermi energy ($E_F$) of $-0.27$~eV with respect to the valence band maximum. Shaded planes at $k_z=0$ and $-\frac{\pi}{a}$ indicate the positions of extracted $k_x$-$k_y$ contours in (c)-(d) respectively. (c)-(d) Fermi surface contours obtained with ARPES at $E_F\pm25$~meV collected using h$\nu=135$ and 110~eV photons to approximate the $k_z=0$ and $k_z=\pi/a$ contours respectively, and simulated DFT images for the Fermi surface in (b). The simulated DFT ARPES images were produced with a set of 600 planes along $k_z$ between $\pm 2.4$~ \AA$^{-1}$, centred on the plane of interest. The contour was generated from the intersection of each plane with the Fermi surface and is realistically blurred to simulate the finite effective energy and $k_x$-$k_y$ resolutions in experiment.  Finally all planes were summed up and weighted with a Lorentzian to simulate $k_z$ broadening.
The $R3m$ Brillouin zone boundary is overlaid as a guide and high-symmetry points are labelled.   }
\end{figure} 
\indent ARPES measurements, performed with photon energies of $h\nu=135$ and $110$~eV, chosen to approximately probe planes which pass through the centre and the $L/Z$-points of the Brillouin zone respectively, are shown in Figure 2. The measured band dispersions along $X/P$-$\Gamma$-$X/P$ directions (Figure 2(a)) show the three Te $p$-derived bands found from DFT. Although the measurements were performed at 15 K, in the rhombohedral phase, it is not possible to identify signatures of the band splittings which result from the breaking of cubic symmetry or the small differences between the dispersions along the $\Gamma$-$X$ and $\Gamma$-$P$ directions.  This is likely predominantly due to large $k_z$ broadening in our measurements, which is a result of the surface sensitivity of photoemission. \\
\indent Nonetheless, comparison with the calculations is instructive. In particular, we find that the valence bands just cut the Fermi level along the $\Gamma$-$X/P$ direction. This indicates a doping level such that the $L$-points are connected by tubular Fermi surfaces  (Figure~2(b)). Consistent with this, we show in Fig. 2(c) and (d) ARPES Fermi surface measurements corresponding to the planes shown in Figure 2(b). These are compared with simulated spectra from our DFT calculations, performed for a Fermi level $0.27$~eV below the valence band maximum (corresponding to a calculated hole density $n_h=2.8\times10^{20}$~cm$^{-3}$, close to the  value from Hall effect measurements in Table I). The bands forming the tube-like regions are directly visible in a plane encompassing the $L$-points of the Brillouin zone (Fig. 2(d)), while the cross-sections of these tubular sections is apparent in the Fermi surface measured in the $k_z=0$ plane (Fig. 2(c)). \\
\indent  Resistivity for  sample S2 normalised to the value at 300~K $(\rho_{300\textrm{\scriptsize~K}})$ is shown in Figure 3(a). An  anomaly due to the increased electron-soft phonon scattering exists at $T_c$ \cite{Katayama},  made clearer by differentiation ($T_c$ and $n_h$ are listed in Table~I).   The magnetoresistance, MR = ($\rho(B) - \rho(0))/\rho(0)$, up to $B=15$~T at 40~mK for a series of applied field angles, $\theta$,  is shown in Figure 3(b).   Here $\theta=0$ corresponds to  $B\parallel$ [001]. Rotation was in increments of $5^\circ$ to $B\parallel$ [011] at $\theta=45^\circ$ through to the $B\parallel$  [010] axis at $\theta=90^\circ$.    Shubnikov-de Haas quantum oscillations (SdH) exist above 8~T and were extracted by subtracting  a smoothly varying polynomial background. Examples of the SdH  against  $B^{-1}$ for  $\theta =0^\circ$ and $\theta=45^\circ$ are shown in  Figure 3(c) and  contain a beating pattern characteristic of several neighbouring frequencies.  Fast Fourier transforms (FFT) resolved 3-4 distinct oscillation frequencies that are labelled $\beta_1$-$\beta_4$, in Figure 3(d). The angular dependence of the oscillation frequencies, $F$,  is shown in Figure 3(e).   The lowest frequency, $\beta_1$, follows
\begin{equation}
F(\gamma) = \frac{F(0)}{\sqrt{\textrm{cos}^2\gamma + \frac{1}{\epsilon} \textrm{sin}^2\gamma}} 
\end{equation}
expected for an ellipsoid with eccentricity, $\epsilon$, shown as the black dashed line.  Here $\gamma$ is the angle with respect to the principal axis of the ellipsoid ((1,1,1) direction).  Frequencies $\beta_2$-$\beta_4$ clearly contain more structure than can be explained by a simple ellipsoid.  \\
\begin{figure}[t]
\includegraphics[scale=0.46]{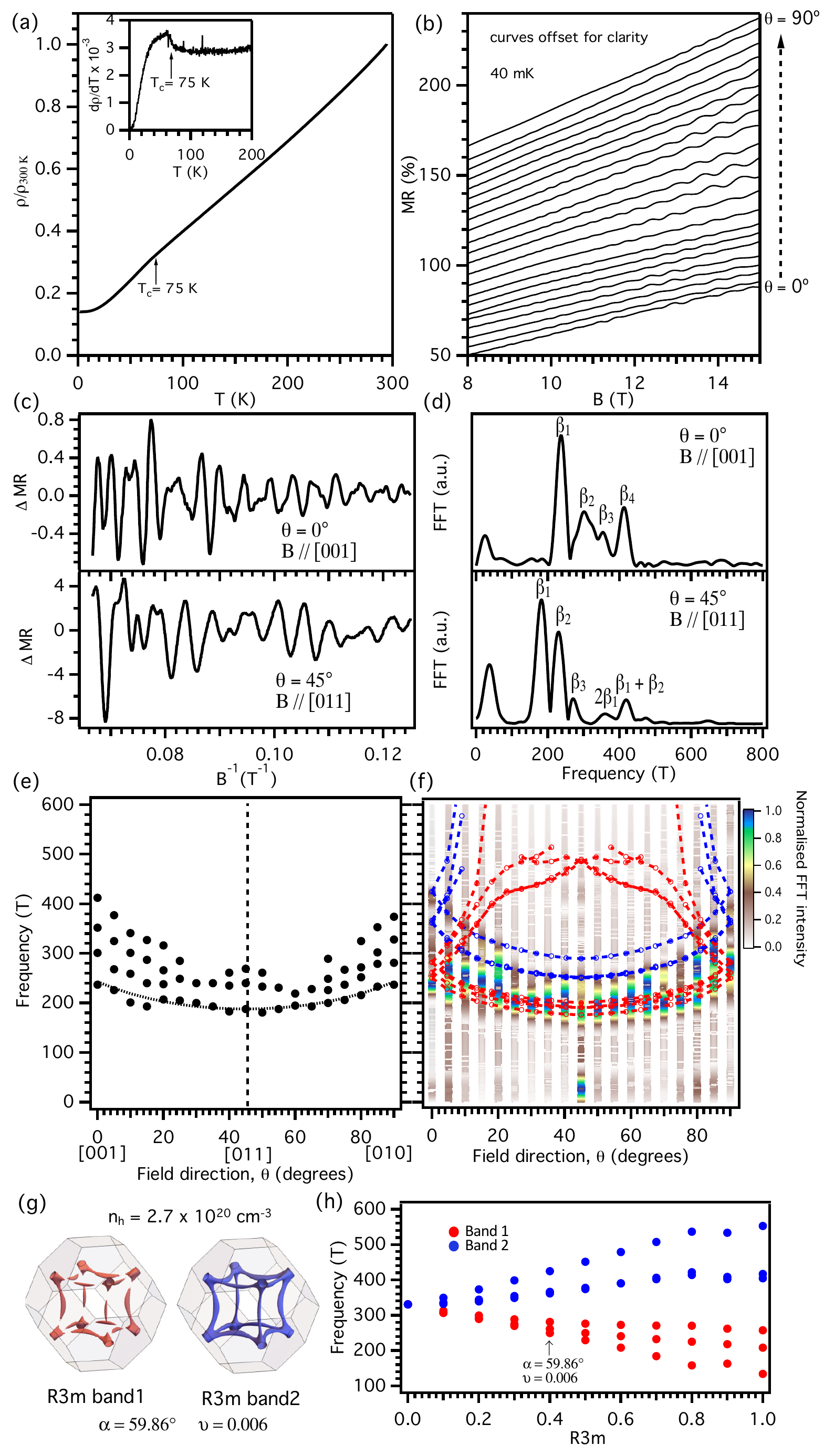}
\caption{\footnotesize  (a) The resistivity, $\rho$, normalised to the value at  300 K  for sample S2.  The anomaly at $T_c =75$ K is  made clearer by differentiation (inset). (b)  The magnetoresistance, MR = ($\rho(B) - \rho(0))/\rho(0)$, at 40 mK for a series of applied field angles, $\theta$. Curves have been offset for clarity. Above 8~T Shubnikov-de Haas  oscillations exist.  (c)~Examples of the  oscillations  plotted against inverse field  for $\theta=0^\circ$ ($B\parallel [001]$) and $\theta=45^\circ$ ($B\parallel [011]$). (d)~Fast Fourier transforms (FFT) of  (c), where  3-4 distinct frequencies labelled $\beta_1$-$\beta_4$ are resolved. (e) The evolution  of $\beta_1$-$\beta_4$, shown as markers, as a function of $\theta$. The black dashed line corresponds to an ellipse given by EQ~(1) with an eccentricity of $\epsilon=5$. (f)~A colourscale of FFT amplitude normalised to the maximum value as a function of field angle and  frequency. Red (band 1) and blue (band 2) markers with dashed lines are the expected oscillation frequencies for the pockets at $L$ in the $R3m$ calculated Fermi surfaces in (g) where $n_h=2.7\times10^{20}$~cm$^{-3}$ and the distortion is $\alpha=59.86^\circ$ and  $\upsilon=0.006$.   (h) Shows the calculated oscillation frequencies for the pockets at $L$ with $B\parallel[001]$ as a function of the distortion size.  Here 0 is the $Fm\overline{3}m$ structure and 1 the the R3m distortion that minimises the DFT energy. The actual distortion is taken to be the value at the arrow.      }
\end{figure} 
\begin{figure}[t]
\includegraphics[scale=0.32]{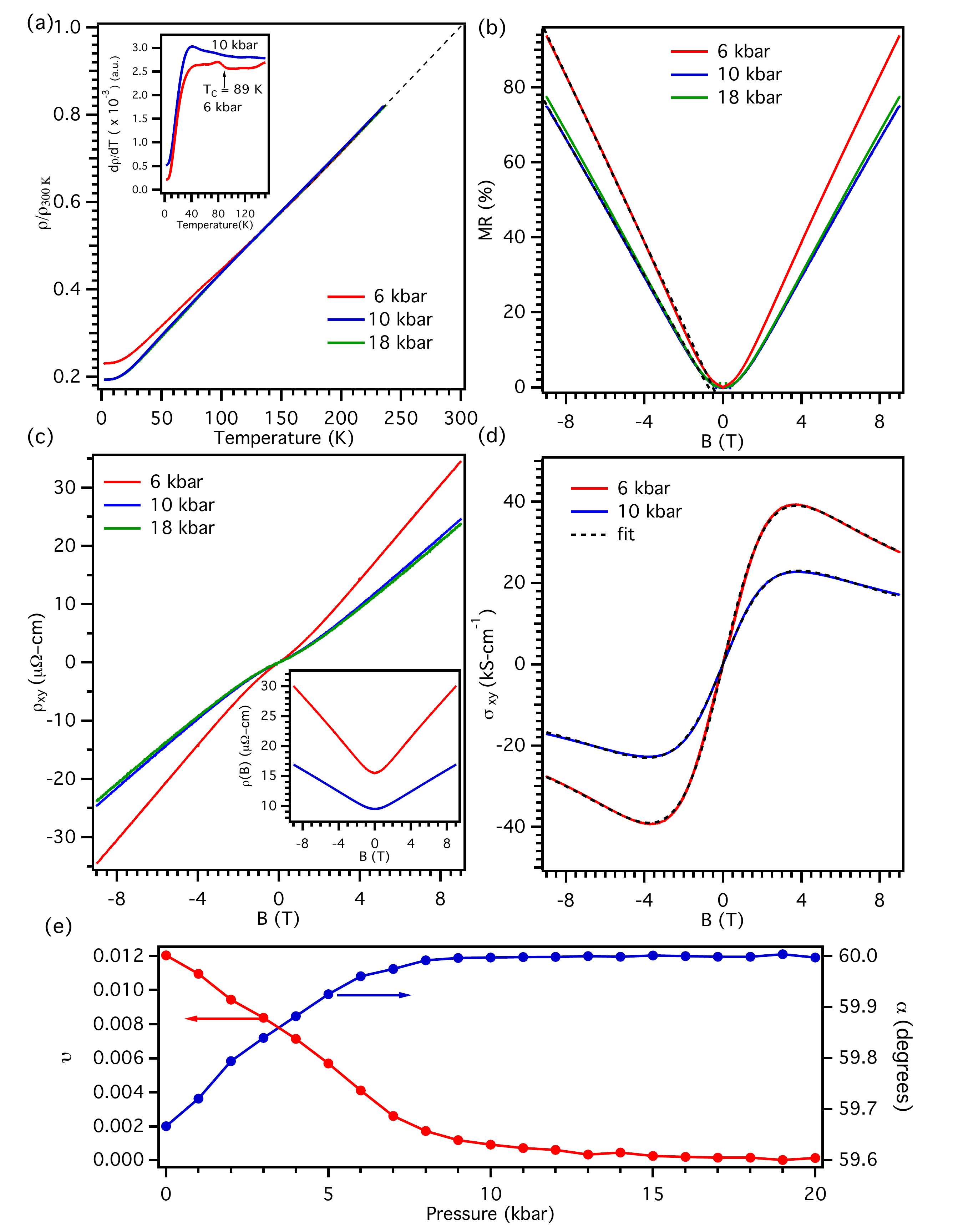}
\caption{\footnotesize   (a)~The resistivity, $\rho$, of sample S3 normalised to the value at 300 K, for pressures   6, 10 and 18~kbar. Here the high temperature freezing of Daphne oil  is avoided by extrapolation (dashed line). Inset $d\rho/dT$ gives $T_c=89$~K at 6~kbar. No anomaly due to $T_c$ can be identified at 10~kbar. (b) Magnetoresistance at 2~K with $B\parallel [001]$ for the above pressures.  The black dashed lines are straight line fits between 1-9~T. (c)  The Hall resistivity, $\rho_{xy}$. Inset are the corresponding curves of $\rho(B)$ with no normalisation. (d) The Hall conductivity, $\sigma_{xy}$, at 6 and 10~kbar. Fits to EQ~(5) are shown as black dashed lines. (e) Pressure dependence of  polar displacement, $ \upsilon$, (left axis) and the rhombohedral angle, $\alpha$, (right axis)  as function of pressure, from spin-orbit coupling DFT calculations. }
\end{figure} 
\indent A  contour plot of the FFT spectra normalised to the largest peak amplitude as a function of frequency and angle is shown in Figure 3(f). Also included are the calculated frequencies for the $L$-pockets of both $R3m$  Fermi surface bands  with $n_h=2.7\times10^{20}$~cm$^{-3}$  and a distortion of $\alpha=59.86^\circ$ and  $\upsilon=0.006$.  These values of $n_h$,  $\alpha$ and  $\upsilon$  give excellent agreement between experimental results and calculation in Figure 3(f).    The Fermi surfaces  are shown in Figure 3(g).      The  $\alpha$ and $\upsilon$ values were determined by how the oscillation frequencies vary as a function of  the rhombohedral distortion. For this we constructed a set of linearly interpolated crystal structures between the cubic $Fm\overline{3}m$ and optimized  $R3m$ structures ($\alpha=59.66^\circ$ and  $\upsilon=0.012$). The change in frequencies for the $L$ pockets with $B\parallel[001]$  as a function of distortion is  in Figure 3(h). This illustrates the substantial effect of the rhombohedral distortion on Fermi surface topology.      \\
\begin{figure}[t]
\includegraphics[scale=0.41]{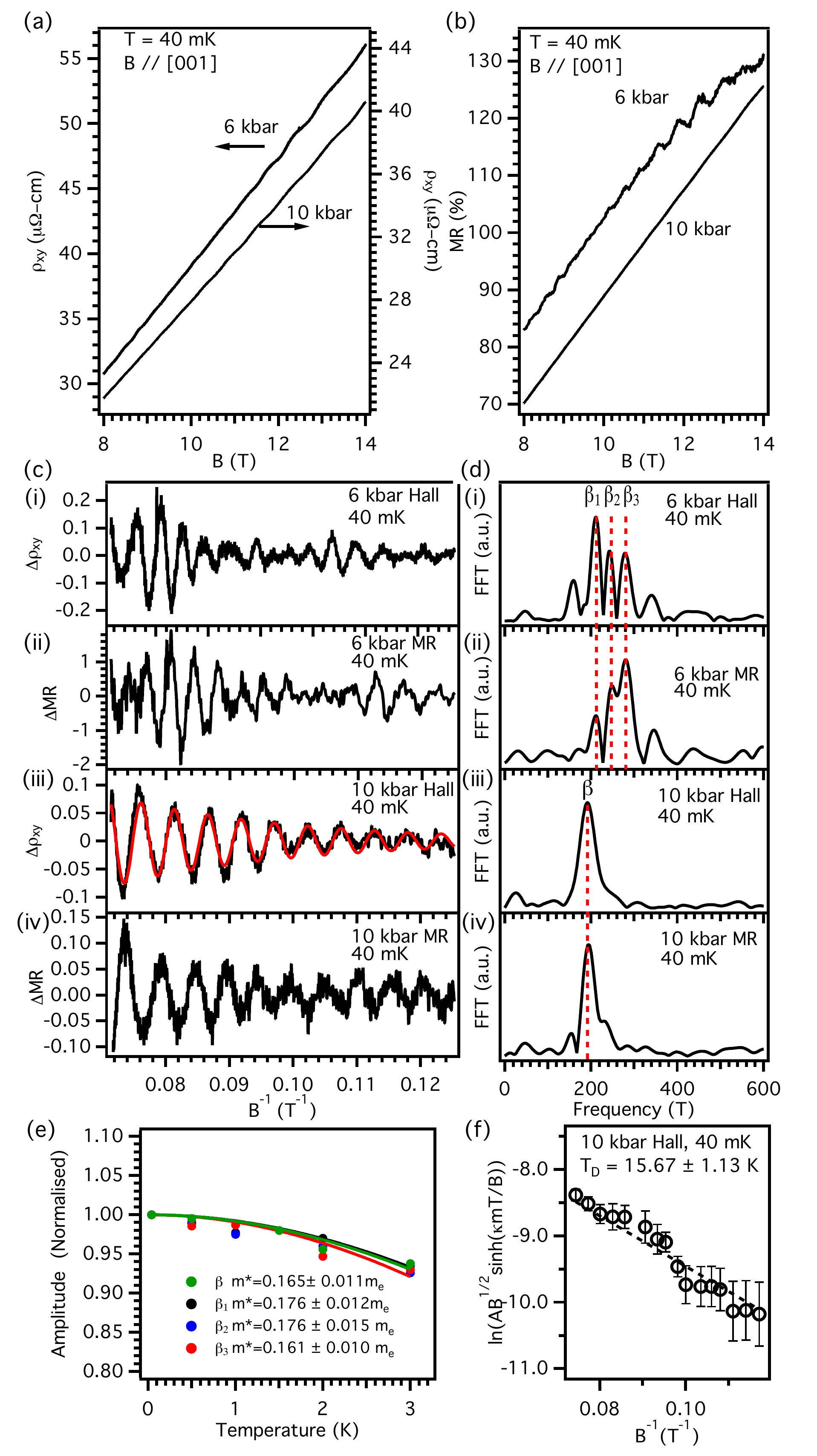}
\caption{\footnotesize (a) The Hall resistivity, $\rho_{xy}$, at 40 mK with $B\parallel$  [001] for 6~kbar (left axis) and 10~kbar (right axis). (b) The corresponding magnetoresistance (MR) at 6 and 10~kbar.  Shubnikov-de Haas oscillations exist above 8~T in both $\rho_{xy}$ and the MR. (c) The oscillations against $B^{-1}$ in; (i) $\rho_{xy}$ (ii) the MR for 6 kbar and  (iii) $\rho_{xy}$ (iv) the MR for 10 kbar. (d);(i)-(iv) The associated fast Fourier transforms of the  oscillations in (c).  At 6 kbar in (i)-(ii) three  frequencies marked $\beta_1$-$\beta_3$ exist in both $\rho_{xy}$ and the MR. A single frequency, $\beta =198$~T, exists for 10~kbar in (iii)-(iv).    (e) The oscillation amplitude in $\rho_{xy}$  against temperature, normalised to the value at 40~mK, for $\beta_1$-$\beta_3$ at 6 kbar and $\beta$ at 10 kbar. Solid lines are fits to EQ~(3) that give the calculated effective masses given in the legend. (f)  The Dingle plot  of the $\beta$ oscillation amplitude at 10~kbar  in $\rho_{xy}$ at 40~mK.  Here a straight line fit gives $T_D = 15.7\pm1.1$~K. For verification the solid red line in (c)(iii) corresponds to an oscillation with frequency 198~T  and amplitude given by EQ~(4) with  $T_D = 15.7$~K.     }
\end{figure} 
\indent Resistivity curves normalised to  $\rho_{300\textrm{\scriptsize~K}}$  for sample S3 studied under pressure are in Figure~4(a) with $d\rho/dT$ inset.  Here $\rho_{300\textrm{\scriptsize~K}}$ was determined by extrapolation (black dashed line) to avoid a high temperature anomaly where the Daphne oil freezes. At 6~kbar  $d\rho/dT$ gives $T_c = 89$~K. On increasing pressure to 10~kbar no  anomaly in $d\rho/dT$ can be identified.  There is also a sharp drop in  $\rho/$$\rho_{300\textrm{\scriptsize~K}}$ by a factor of 1.58 at 2~K. No further changes were observed at 18~kbar and the curve is almost indistinguishable from that at 10~kbar. The magnetoresistance  and  Hall resistivity, $\rho_{xy}$,  at 2 K are shown in Figure 4(b) and (c) respectively. The Hall conductivity, $\sigma_{xy}$,  is given by
\begin{equation}
\sigma_{xy}= \frac{\rho_{xy}}{\{\rho(B)\}^2 + \rho_{xy}^2}
\end{equation} 
and  is shown in Figure 4(d).\\
\indent The Hall resistivity and magnetoresistance  measured on sample S3 up  to 14~T at 40~mK are shown in Figure 5(a) and (b).  Both quantities contain Shubnikov-de Haas oscillations (SdH) above 8~T. Here, only measurements with $B\parallel [001]$ could be taken.    Figure 5(c) shows the  SdH against $B^{-1}$ for 6~kbar in (i)~the Hall effect and (ii)~the MR and for 10~kbar in (iii)~the Hall effect and (iv)~the MR.  The FFT  in Figure 5(d)  shows  three frequencies exist in both the Hall and  MR at 6 kbar; $\beta_1 = 209$~T, $\beta_2 = 246$~T and $\beta_3=290$~T.  Only a single frequency in the FFT at $\beta = 198$ T exists at 10~kbar in Figure 5(d) for both the Hall and  MR. \\
\indent The temperature dependence of the quantum oscillation  amplitude is governed by the temperature reduction term, $R_T$, in the Lifshitz-Kosevich formula given by
\begin{equation} 
R_T = \frac{\kappa m^* T/B}{sinh(\kappa m^* T/B)}
\end{equation}
where $\kappa=2\pi^2k_B/e\hbar$ and $m^*$ is the effective mass. The SdH amplitude normalised to the value at 40 mK  with corresponding fits to $R_T$  are shown in Figure 5(e). Measurements were limited to 3 K, the maximum temperature of the dilution refrigerator, where only a slight decrease in amplitude could be realised.   Derived effective masses for $\beta_1$-$\beta_3$  are the same within error with an average value of $m^*=0.171\pm0.012$~$m_e$.  At 10~kbar no change in $m^*$ was seen within error with $\beta$ and $m^*=0.165\pm0.011$~$m_e$. \\
\indent The quantum oscillation amplitude dependence on field is determined by the Dingle term, $R_D$, in the Lifshitz-Kosevich formula given by
\begin{equation}
R_D = \textrm{exp}\left(\frac{-\kappa m^*T_D}{B}\right)
\end{equation}
where the  Dingle temperature $T_D = \hbar/2\pi k_B \tau$ can be used to calculate the scattering time $\tau$. Beating patterns  at 6~kbar make any estimate of $T_D$ highly uncertain. At 10 kbar a graph of $\textrm{ln}(AB^{1/2}\textrm{sinh}(\kappa m^*T/B))$ against $B^{-1}$ is shown in Figure 5(f), where $A$ is the oscillation amplitude. A straight line fit gives $T_D = 15.7\pm1.1$~K and consequently $\tau = 7.76 \pm 0.55 \times 10^{-14}$~s. The solid red line in Figure 5 (c) is a calculation with $T_D =15.7$ K for a single oscillation frequency  of $198$~T, showing excellent agreement with the measurements.  \\
\indent Having presented our theoretical and experimental findings we now discuss the significance of the results. DFT bandstructure calculations highlight the large changes in Fermi surface topology as a consequence of the polar distortion.  They also reproduce a small widening of the bandgap and shifts in the density of states, seen previously with ARPES \cite{Littlewood_Arpes}.     The qualitative agreement with our ARPES measurements reported here provides a validation in the accuracy of our DFT calculations close to the Fermi surface and confirms that the off-stoichiometry leads to a degenerate doping i.e.\ the Fermi level shifts into the valence bands.\\
\indent Quantum oscillations below $T_c$ contain 3-4 neighbouring frequencies.  The DFT calculation identifies the detected oscillations to be from the $L$-pockets that have a similar angular dispersion, even when  these pockets are connected by tube like Fermi surfaces. The expected oscillation frequencies for the connecting tubes are all less than 80~T for the Fermi surface in Figure 3(g) and were not resolved in experiment.  Although peaks exist at $\approx 30$~T in the FFT spectra,  they may also be a result of subtracting the polynomial background. As there are  four unique $L$-pockets that vary in size in the rhombohedral phase  the presence of several neighbouring oscillation frequencies for $B\parallel$ [001] is expected in the $R3m$ structure (Figure 3(h)).     In the cubic phase  the single eight-fold degenerate $L$-pockets are expected to give only a single oscillation  frequency  for $B\parallel$ [001].  With pressure we find 10~kbar is adequate to merge the neighbouring frequencies into a single frequency. Additionally at 10~kbar no anomaly in resistivity due to $T_c$ can be identified.  Therefore  the rhombohedral transition has been suppressed by a pressure of 10~kbar.  This is also accurately captured in DFT calculation, shown in Figure 4(e) for  $\upsilon$ and $\alpha$ as a function of pressure.      \\
\indent The suppression of the rhombohedral transition leads to large changes in Hall signal. $\rho_{xy}$ shows  non-linear behaviour for all pressures. Our measurements for $\sigma_{xy}$ at all pressures are well described by a single band model,
\begin{equation}
\sigma_{xy} =eB\left( \frac{n_h\mu_h^2}{1+\mu_h^2B^2} \right) = \frac{\sigma\omega_c\tau}{1+\omega_c^2\tau^2}
\end{equation}
where $\mu_h$, is the carrier mobility, $\sigma$ the zero field conductivity and $\omega_c$ the cyclotron frequency.  EQ(5) highlights that for sufficiently mobile carriers with $\mu_hB\sim 1$  non-linearity may still exist for a single band.  Using a two-band model did not improve the fit despite adding two extra fitting parameters.  Values of $n_h$ and $\mu_h$ extracted from the fits  are given in Table~II.  Upon entering the cubic phase the carrier concentration increases significantly by a factor of 1.64, while the mobility stays approximately constant within error, consistent with the drop  in resistivity by a factor of 1.58.    \\
 \begin{center}
\begin{tabular}{|c|c|c|}
\hline
   & $R3m$ (6 kbar)   & $Fm\overline{3}m$ (10 kbar)\\ \hline 
$n_h$  ($\times 10^{20}$ cm$^{-3} $)  & $1.12\pm0.04$  &$1.83\pm0.05$\\\hline
$\mu_h$  (cm$^2$ V$^{-1}$s$^{-1}$) & $2570\pm50$&$2660\pm50$ \\\hline
\end{tabular}  
\end{center}
\noindent Table II : {\footnotesize The  values of carrier concentration, $n_h$, and mobility, $\mu_h$, in sample S3 according to the fits of $\sigma_{xy}$ in Figure 4(d).}\\

 \indent On increasing pressure in the cubic structure (to 18 kbar) no further changes  are observed in any measured quantity. The changes in  $\rho_{xy}$ and $\sigma_{xy}$ are therefore a direct consequence of the modification in Fermi surface associated with the distortion.  Expected quantum oscillation frequencies from calculations show the distortion acts to increase the volume of some $L$-pockets while decreasing others  (Figure 3(h)).  This is not seen in experiment. Figure 5(d) shows $\beta_1$-$\beta_3$ at 6~kbar are all larger in magnitude than $\beta$  at 10~kbar.  A full angular dependence  is required  to say definitively how the pockets change. However these results point towards the $L$-pockets being smaller in the cubic phase with potentially more carriers in the connecting tubes  to maintain a fixed total number of carriers. \\
 \indent Quantum oscillations   at 10 kbar give   $\mu_h =e\tau/m^*= 846\pm59$~cm$^2$V$^{-1}$s$^{-1}$, significantly less than the mobility from the Hall effect  in  Table II.   This suggests that the Hall conductivity is dominated by the connecting tubes that have a  significantly higher mobility than the $L$-pockets.  Therefore for the cubic phase more carriers in the connecting tubes in turn leads to more carriers becoming visible to the Hall effect, providing a simple explanation for  the changes seen here. Since the Dirac points are not close to  $E_F$ the role of  surface contributions can be discounted. \\
\indent  The MR in Figure 4 (b) follows a similar behaviour in both structures. It begins quadratically  for low $B$ before becoming linear above 1 T to the highest fields, demonstrated by the straight line fits.  Linear MR is an unusual phenomenon and both quantum and classical explanations have been proposed. Abrikosov's well known quantum model gives a linear MR when all carriers occupy the lowest Landau level for a linear electron energy dispersion \cite{Abrikosov_1, Abrikosov_2}.  The linear MR reported in some topological insulators is expected to be associated with similar quantum effects \cite{Wang_2012}.  Previous reports on SnTe thin films with a larger $n_h$ than our samples, saw a linear magnetoresistance and attributed it to Dirac surface states dominating the transport as a result of band bending at the substrate/film interface that brings $E_F$ closer to the Dirac points \cite{Wei}. In our bulk crystals  such band bending is unlikely. \\
\indent  Classically linear MR from irregular current paths  as a result of  disorder was described by Parish and Littlewood~\cite{Parish}.  Here our samples have high mobility carriers and quantum oscillations that reflects relatively  little disorder, making a Parish Littlewood mechanism unlikely. Instead we suggest a straightforward origin first discussed by Pippard based on a  Fermi surface containing sharp corners   \cite{Pippard}.  \\
\indent Pippard demonstrated that for sufficiently small values of $\omega_c\tau\ll2\pi$   Fermi surfaces with sharp corners   lead to a MR varying as $\omega_c\tau$ rather than $\omega_c^2\tau^2$  \cite{Pippard}.  While the occurrence of infinitely sharp corners is difficult to imagine in reality, Pippard points out regions with suitably  sharp curvature and minimal rounding  lead to an initial quadratic form at low field followed by linear behaviour with increasing field, analogous to the measurements reported here.         At 9 T in the cubic phase  $\omega_c\tau\approx0.6$, satisfying the requirement $\omega_c\tau\ll2\pi$.  We therefore attribute the linear MR in SnTe to regions of high curvature present in the Fermi surface for both the cubic and rhombohedral structure, most likely  at  the necks where $L$-pockets and  connecting tubes meet making an almost right angle.   Curvature driven linear MR has previously been reported in materials with partially gapped Fermi surfaces due to density wave order~ \cite{Kikugawa_2010, Feng}, proving it as a potential explanation for many other systems.\\
\indent In conclusion we have shown the polar distortion in SnTe is accompanied by a reconstruction of the Fermi surface. Under ambient pressure both our quantum oscillation and ARPES measurements are well described with DFT calculations.     Also captured by  DFT  is a small widening of the band gap and shift in density of states at the Fermi level for the polar phase, previously seen with ARPES.  We found that  a pressure of just under 10~kbar stablizes the cubic structure in good agreement with the required pressure calculated with DFT. This  consequently  allows the contribution to the Hall effect for each feature of the Fermi surface to be determined.   Linear magnetoresistance  has been observed in both structures and attributed to the presence of regions with high curvature in the Fermi surface.    \\
{\bf{Acknowledgements}}; We wish to gratefully thank support from the UK EPSRC grants EP/P013686/1 and EP/R013004/1 (CDON and ADH) and the Royal Society (PDCK) and the Leverhulme Trust (PDCK and FM).  We also acknowledge  PhD studentship support  from ESPRC EP/L015110/1 (HDJK) and EP/K503162/1 (OJC) and via the International Max-Planck Research School for Chemistry and Physics of Quantum Materials (IM).  Computational resources provided by the UK's National Supercomputer Service through the UK Car-Parrinello consortium (No. EP/P022561/1) and by the UK Materials and Molecular Modelling Hub (No. EP/P020194) are gratefully acknowledged. Access to the CASSIOPEE beamline (proposal number 20170362) at SOLEIL is also gratefully acknowledged.

\end{document}